# Circular symmetric Airy beam with the inverse propagation of the abruptly autofocusing Airy beam

CHUANGJIE XU,[1, *]

[1]*Room 832, Apartment 353, Sun Yat-sen University, Guangzhou 510275, China*
*Corresponding author: xuchj7@mail2.sysu.edu.cn*

**In this letter, we introduce a new class of light beam, the circular symmetric Airy beam (CSAB), which arises from the extensions of the one dimensional (1D) spectrum of Airy beam from rectangular coordinates to cylindrical ones. The CSAB propagates at initial stages with a single central lobe that autofocuses and then defocuses into the multi-rings structure. Then, these multi-rings perform the outward accelerations during the propagation. That means the CSAB has the inverse propagation of the abruptly autofocusing Airy beam. Besides, the propagation features of the circular symmetric Airy vortex beam (CSAVB) also have been investigated in detail. Our results offer a complementary tool with respect to the abruptly autofocusing Airy beam for practical applications.**

http://dx.doi.org/10.1364/OL.99.099999

Since the Airy beams with the diffraction free, self-healing and self-acceleration properties have been proposed in 2007 [1, 2], they have become a subject of immense interest in the last decade [3–9]. Among research topics, the generation [2, 10], the applications [11, 12, and 13] and control [5, 14, 15] of the Airy beams have attracted much attention recently. Besides, the circular Airy beam (CAB) with the abruptly autofocusing property has been proposed by extending the Airy function from rectangular coordinates to cylindrical ones [16]. These beams abruptly focus their energy right before the focal point while maintaining a low intensity profile until that very point. This property enables the CAB focus at the target while before that it doesn't cause damage. Due to its own features, the CAB possesses wide applications in biomedical treatment or optical micromanipulation [16, 17, 18].

Otherwise, a conventional Airy beam with rectangular symmetry called symmetric Airy beam (SAB) which is arisen from the symmetrization of the spectral cubic phase of an Airy beam has been presented [19, 20 and 21]. The SAB propagates at initial stages with a single central lobe that autofocuses and then collapses into two specular off-axis parabolic lobes like those corresponding to two Airy beams accelerating in opposite directions. This 'autofocus to defocus' process is dynamically inverse to what happens with CAB, in addition to their different spatial symmetries, indicating that both beams may be seen as complementary for practical applications. What's more, slight misalignment of the vortex and the SAB enables the guiding of the vortex into one of the self-accelerating channels of the main lobes [22]. However, a beam not only having the inverse propagation process of CAB but also possessing circular symmetry has not been proposed ever. Does this kind of beam exist? The answer is sure.

In this Letter we develop a formalism to generate an Airy beam having circular symmetry, autofocusing properties and the inverse propagation of CAB. Mathematically, this circular symmetric Airy beam arises from a finite-energy Airy beam by only replacing the 1D spatial frequency with the radial spatial frequency in its spectrum. We perform a numerical analysis showing its main properties. Besides, the influence of the different parameters on the autofocusing properties has been investigated. At last, the propagation features of the circular symmetric Airy vortex beam (CSAVB) also have been presented in detail.

At the initial stage of propagation $z = 0$, the Fourier transform of the spatial field $u\left(\frac{x}{w_0}, 0\right) = Ai\left(\frac{x-K_0}{w_0}\right) exp\left[a\left(\frac{x-K_0}{w_0}\right)\right]$ ($K_0$ is the translation factor and $a$ is the decay factor) is explicitly given by:

$$\tilde{u}(k,0) = w_0 e^{a^3/3} e^{-ak_x^2 w_0^2 + ik_x^3 w_0^3/3 + ik_x w_0 a^2} e^{-ik_x w_0 K_0}, \quad (1)$$

where $k_x$ is the 1D spatial frequency. Then we perform the variable substitution by replacing the 1D spatial frequency with the radial spatial frequency and we can obtain the initial spectrum of the CSAB:

$$\widetilde{U}(K,0) = w_0 e^{a^3/3} e^{-aw_0^2 K^2 + iw_0^3 K^3/3 + iKw_0 a^2} e^{-iw_0 KK_0}, \quad (2)$$

where $K = \sqrt{k_x^2 + k_y^2}$ is the radial spatial frequency and $k_i$ ($i = x, y$) are the 1D spatial frequencies. The wrapped spectral phase mask for the resulting CSAB when the variable substitution is carried out is shown in Fig. 1.

An analytic expression for a CSAB was not found. However, its paraxial solution $U(r, z)$ can easily be built by using the angular spectrum formalism [23]

$$U(r,z) = 2\pi \int_0^\infty \widetilde{U}(K,0) \exp\left(ikz\sqrt{1-\lambda^2 K^2}\right) J_0(2\pi rK) K dK, \quad (3)$$

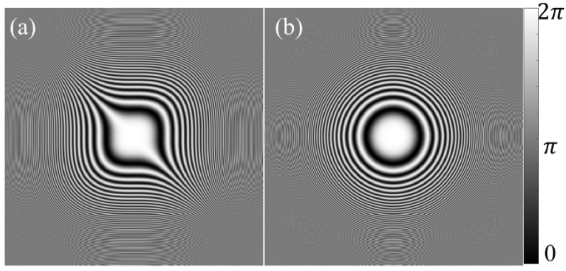

Fig. 1 The wrapped spectral phase masks for a conventional Airy beam and for the resulting CSAB when we perform the variable substitution; the phase ranges correspond to $(-1,1)\,mm$.

where $k = \frac{2\pi}{\lambda}$ is the wave number. By use of the Eq. (3), we can simulate the propagation of the CSAB numerically. In this paper, we set the parameters $\lambda = 589.0\,nm$, $K_0 = 1\,mm$, $a = 0.10$, $w_0 = 1\,mm$ and the Rayleigh distance $Z_R = kw_0^2$. We use the Rayleigh distance to normalize the propagation distance. Here, we definite $I_0$ as the maximum intensity of CSAB at the initial plane. In this letter, the intensity plots have been normalized by the $I_0$ except the maximum intensity plots.

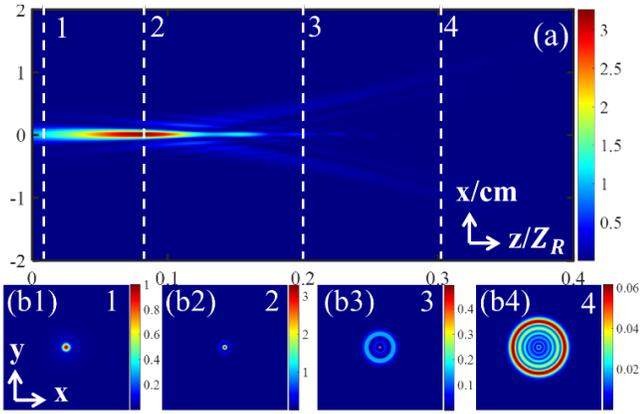

Fig. 2 The propagation of the CSAB in the free space; (a) side view of CSAB propagation numerically; (b1) - (b4) snapshots of transverse intensity patterns of CSAB at planes 1 - 4 marked in (a).

At the beginning, we would like to discuss the propagation properties of the CSAB in the free space. Figure 2 plots the side view and the transverse intensity patterns at different planes of CSAB in which we can learn that the CSAB first performs the autofocusing and then transforms into the multi rings structure performing the outward acceleration. But after the autofocusing plane (plane 2) most of the energy first concentrates on the center of the intensity pattern and then gradually transfers to the outside ring as we can see in the Fig. 2(b3) - 2(b4). This 'autofocus to defocus' process just acts as the inverse propagation process of the CAB and the CSAB has the circular symmetric distribution. That means both beams may be seen as complementary for practical applications to each other. What's more, this new kind of autofocusing beam may also be useful in biological treatment, microscopic imaging and optical micromanipulations.

Next, we would like to discuss the influence of the decay factor $a$ on the propagation of the CSAB. Figure 3 depicts the propagations and Fig. 4 plots the maximum intensity of the CSAB in the free space with different decay factors $a = 0.15, 0.20, 0.25$, respectively. We can find that the decay factor greatly affects the focal length and the focal intensity of the autofocusing beams. Otherwise, we are able to reduce the focal intensity and shorten the focal length of CSAB by increasing the decay factor which makes the autofocused beam more flexible. But the change of the decay factor doesn't affect the size of the focus spot in the autofocusing plane. More interestingly, we deduce that the decay factor also could influent the energy distribution from the intensity patterns of Fig. 4(a3) - 4(c3). With the increase of the decay factor, the energy gradually transfers from the center of the intensity pattern to the outward ring at the same plane.

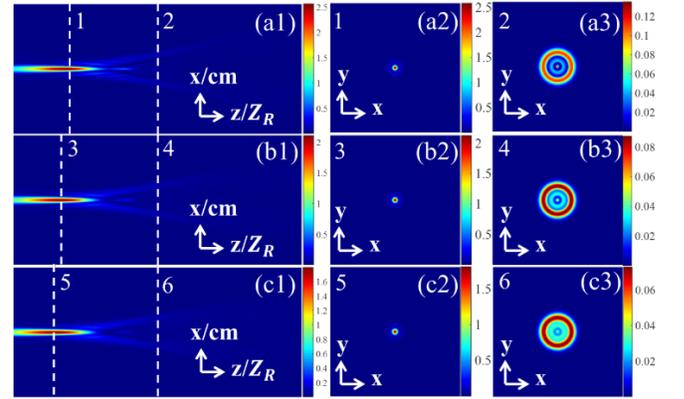

Fig. 3 The propagations of the CSAB in the free space with different decay factors $a$; (a1) - (c1) side views of the CSAB; (a2) - (c2) and (a3) - (c3) snapshots of transverse intensity patterns of CSAB at planes 1 - 6 marked in (a1) - (c1); the decay factors $a$ equal to (a1) - (c3) 0.15, (b2) -(b3) 0.20, (c1) - (c3) 0.25. Other parameters are the same as those in Fig. 2.

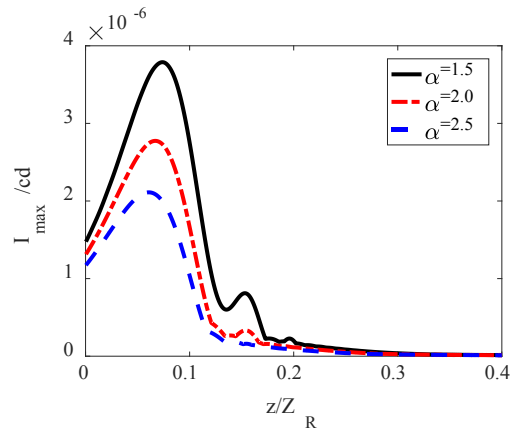

Fig. 4 The maximum intensity plots of CSAB with different decay factors $a$. Other parameters are the same as those in Fig. 2.

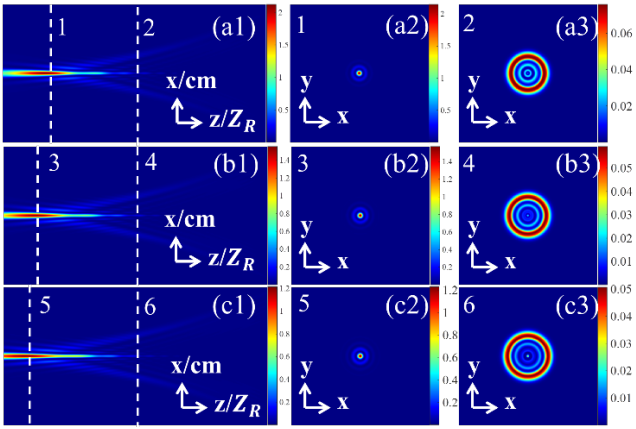

Fig. 5 The propagations of the CSAB in the free space with different translation factors $K_0$; (a1) - (c1) side views of the CSAB; (a2) - (c2) and (a3) - (c3) snapshots of transverse intensity patterns of CSAB at planes 1-6 marked in (a1) - (c1); the translation factors $K_0$ equal to (a1) - (c3) 1.5, (b2) - (b3) 2.0, (c1) - (c3) 2.5. Other parameters are the same as those in Fig. 2.

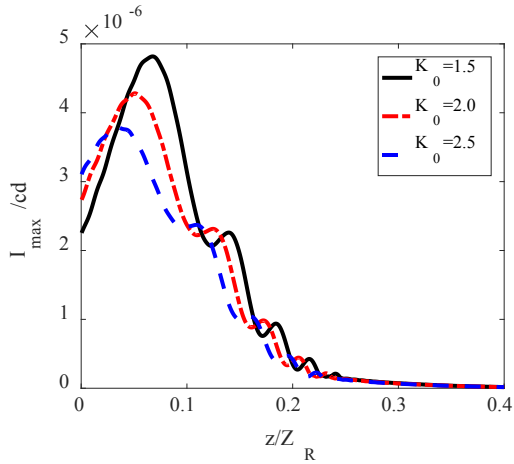

Fig. 6 The maximum intensity plots of CSAB with different translation factors $K_0$. Other parameters are the same as those in Fig. 2.

The value of the translation factor also has a tremendous influence on the propagation characteristics of the CSAB. Figure 5 depicts the propagations and Fig. 6 plots the maximum intensity of the CSAB in the free space with different translation factors $K_0 = 1.5, 2.0, 2.5$, respectively. Similar to the decay factor, the increase of the translation factor would reduce the focal intensity and the focal length of the CSAB. On the other hand, the increase of the translation factor could enlarge the size of the focal spot since it decides the number of the outward rings of the focal intensity pattern as we can see in Figs. 5(a2) - 5(c2). And from Figs. (a3) - (c3), we can learn that the energy would move outside and the outward ring turns larger as the translation factor increasing.

It is well-known that Airy beams have 'self-healing' property. This self-healing property enables the self-regeneration of Airy beam when it encounters obstructions. To check whether the CSAB possesses the self-healing property, we calculated the beam propagation profiles when a radial obstruction object, whose radius is about 1 mm, hinders the CSAB, shown in Fig. 7. From Fig. 7(a), we can learn that the CSAB reconstructs the beam profile before the autofocusing plane quickly when a radial obstruction object hinders it at the initial plane. Figure 7(b) depicts the situation that the obstruction object hinders the CSAB at the autofocusing plane which show that the CSAB not only performs the self-healing property but also focuses again.

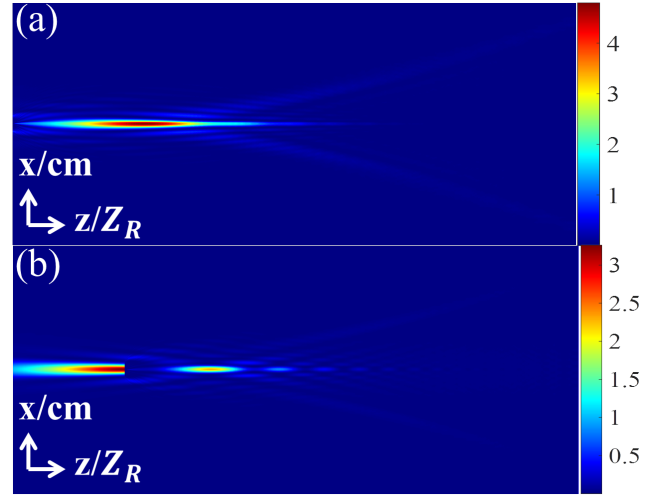

Fig. 7 Numerical results for the self-healing property: obstruction object at (a) initial plane and (b) autofocusing plane. Other parameters are the same as those in Fig. 2.

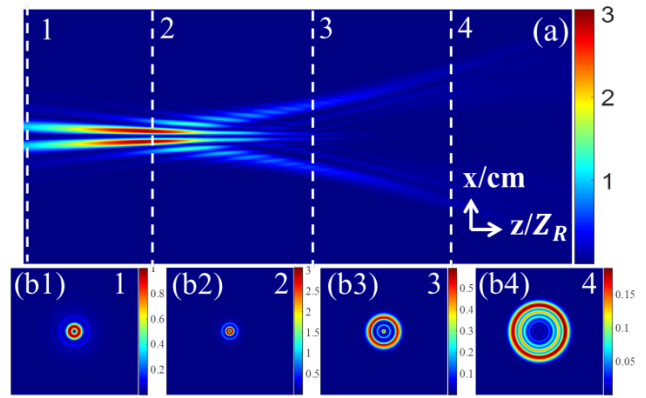

Fig. 8 The propagation of the first order on axis ($l=1$) CSAVB in the free space; (a) side view of CSAVB propagation numerically; (b1) - (b4) snapshots of transverse intensity patterns of CSAVB at planes 1 - 4 marked in (a). Other parameters are the same as those in Fig. 2.

At last, we would pay our attention to investigating the propagation of the CSAVB in the free space. Figure 8 presents the propagation of the CSAVB in the free space. Comparing to Fig. 1, we can learn that the existence of the on axis vortex makes the intensity pattern of the CSAB hollow. The hollow focal spot of the CSAVB enables it to trap the micro particle whose refractive index is lower than that of the surrounding medium. What's more, the vortex also would reduce the energy on the center of the intensity pattern as we can see in Figs. 8(b3) and 8(b4). The influence of the topological charges on the CSAVB has been shown in Fig. 9, from which we can learn that the larger the topological charges are the smaller focal intensity and the focal length are. Besides, the large topological charges would enlarge the size of the focal spot of the CSAVB as shown in Fig. 9(b).

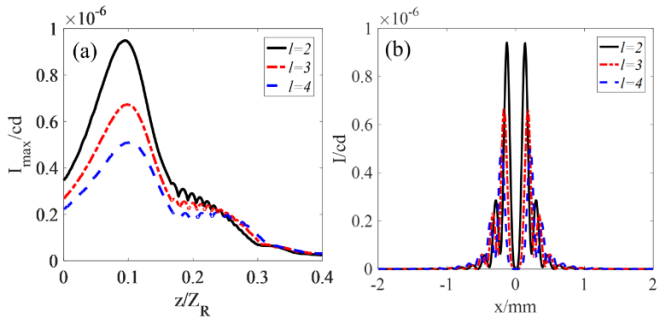

Fig. 9 The maximum intensities plots of CSAB with different topological charges; (a) the maximum intensities of the CSAVB along z direction; (b) the maximum intensities on the focus planes along x direction. Other parameters are the same as those in Fig. 2.

In summary, we propose a new kind of the autofocusing beam arising from the extensions of the 1D spectrum of Airy beam from rectangular coordinates to cylindrical ones. This kind of beam that we call circular symmetric Airy beam (CSAB) first autofocuses into a solid spot and then transforms into multi rings structure performing the outward acceleration during the propagation. This 'autofocus to defocus' process of CSAB is dynamically inverse to what happens with CAB which means that both beams may be seen as complementary for practical applications. And the propagation properties of CSAVB also have been investigated in detail. The hollow focal spot of the CSAVB enables it to trap the micro particle whose refractive index is lower than that of the surrounding medium. What's more, this new kind of autofocusing beam may also be useful in biological treatment, microscopic imaging and optical micromanipulations.

**Disclosures**. We declare that we have no conflict of interest.